\documentstyle[12pt,epsf,epsfig,wrapfig]{article}
\begin{document}
\begin{center}
{\bfseries SMALL-$X$ BEHAVIOUR OF THE NON-SINGLET AND SINGLET STRUCTURE
FUNCTIONS $G_1$.}
\vskip 5mm
\underline{B.I. Ermolaev}$^{1 \dag}$, M. Greco$^{2}$ and
S.I.~Troyan$^{3}$
\vskip 5mm
{\small
(1) {\it Ioffe Physico-Technical Institute, 194021
St.Petersburg, Russia }
\\
(2) {\it Department of Physics and INFN, University Rome III, Rome,
         Italy
}
\\
(3) {\it St.Petersburg Institute of Nuclear Physics, 188300 Gatchina,
         Russia
}
\\
$\dag$ {\it
E-mail: ermolaev@pop.ioffe.rssi.ru
}}
\end{center}
\vskip 5mm

\begin{abstract}
Explicit expressions for the non-singlet and singlet structure functions
$g_1$ at the small $x$-region
are obtained. They include the total resummation of
double-logarithmic contributions and
accounting for the running QCD coupling effects.
We predict  that
asymptotically the singlet $g_1 \sim x^{- \Delta_S}$, with the intercept
$\Delta_S = 0.86$, which is approximately
twice larger than the non-singlet
intercept $\Delta_{NS} = 0.4$. The impact of the initial
quark
and gluon
densities  on the sign of $g_1$ at $x \ll 1$ is discussed.
\end{abstract}

\vskip 8mm

\section{Introduction}

\begin{figure}
\begin{center}
\begin{picture}(280,100)
\put(10,10){
\epsfbox{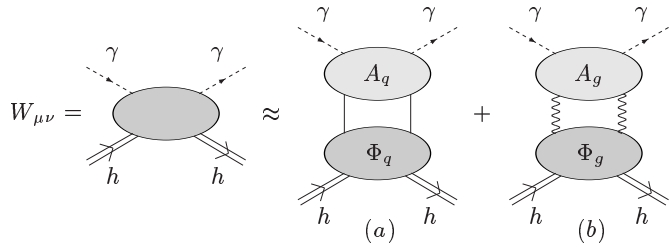}
}
\end{picture}
\end{center}
\caption{ Representation of the hadronic tensor as the covolution of the
fragmentation functions $\Phi_{q,g}$ and the partonic tensor.}
\label{dubnafig1}
\end{figure}

Deep inelastic scattering (DIS)
is one of the basic processes for investigating
the structure of hadrons. As is well known, all information about
the hadrons
participating into DIS comes from the hadronic
tensor $W_{\mu \nu}$. The imaginary part of $W_{\mu \nu}$ is proportional
to the
forward Compton amplitude when the deeply off-shell photon
with virtuality $q^2$ scatters off an on-shell hadron with momentum $p$.
For the electron-hadron DIS, the spin-dependent part,$W_{\mu \nu}^s$, of
$W_{\mu \nu}$ is

\begin{equation}
\label{w}
W_{\mu \nu}^s = \imath \epsilon_{\mu \mu \lambda \rho}
\frac{q_{\lambda} m}{pq} \Big[ S_{\rho} g_1 +
\big(S_{\rho} - \frac{(Sq)}{pq} p_{\rho} \big) g_2 \Big]
\approx \imath \epsilon_{\mu \mu \lambda \rho}
\frac{q_{\lambda} m}{pq} \Big[ S_{\rho}^{||} g_1 +
S_{\rho}^{\perp}\big(g_1 + g_2 \big) \Big]
\end{equation}
where $m$ is the hadron mass,
$S_{\rho}^{||}$ and  $S_{\rho}^{\perp}$ are the longltudinal and
transverse (with respect to the plane formed by $p$ and $q$)
components of the hadron spin  $S_{\rho}$; $g_1$ and $g_2$ are
the spin structure functions. Both of
them depend on $x = - q^2/2pq,~ 0< x \leq 1 $ and $Q^2 = - q^2 > 0$.
Obviously, small $x$ corresponds to $s = (p+q)^2 \approx 2pq \gg Q^2$.
In this case,
$S_{\rho}^{||} \approx  p_{\rho}/m$  and therefore the part of
$W_{\mu \nu}^s$ related to $g_1$ does not depend on $m$. Then if
$Q^2 \gg m^2$, one can assume the factorization and regard  $W_{\mu \nu}^s$
as a convolution of two objects (see Fig.~\ref{dubnafig1}). 
The first of them is
the probability $\Phi$ ($\Phi = \Phi_q$ in Fig.~\ref{dubnafig1}a and 
$ \Phi = \Phi_g$ in
Fig.~\ref{dubnafig1}b) to find a polarized parton (a quark or a gluon) in the
hadron. The second one is the partonic tensor
$\tilde{W}_{\mu \nu}^s$ defined and parametrized
similarly to $W_{\mu \nu}^s$. Whereas
$\Phi_{q, g}$ are essentially non-perturbative objects,
the partonic tensor $\tilde{W}_{\mu \nu}^s$,
i.e. the partonic structure functions $g_1$ and $g_2$, can
be studied within perturbative QCD.
THe lack of knowledge of  $\Phi$ is usually compensated by introducing
initial parton distributions that are found from pheneomenological
considerations. On the contrary, there are regular perturbative
methods for calculating the structure functions in the partonic
tensor $\tilde{W}_{\mu \nu}^s$.
The best known instrument to calculate the structure functions
to all orders in $\alpha_s$ is the DGLAP\cite{dglap} approach.
Once applied to the description of the
experimental data, DGLAP provides good results\cite{a}.
The extrapolation into
the small-$x$ region of DGLAP predicts an asymptotical behaviour
$\sim \exp(\sqrt{C \ln(1/x)\ln\ln Q^2})$ for all DIS
structure functions (with different factors $C$). However, from a
theoretical point of view, such  an extrapolation is
rather doubtful. In particular, it neglects in a systemetical way
contributions $\sim (\alpha_s \ln^2(1/x))^k$ which are small when
$x \sim 1$ but become large when $x \ll 1$. The total resummation of
these double-logarithmic (DL) contributions made in
Refs.~\cite{ber1} and Ref.~\cite{ber} for the non-singlet $(g_1^{NS})$ and
singlet $g_1$ respectively leads to the Regge (power-like) asymptotics
$g_1 (g_1^{NS}) \sim (1/x)^{\Delta^{DL}} ((1/x)^{\Delta^{DL}_{NS}})$,
with $\Delta^{DL}, \Delta^{DL}_{NS}$ being the
intercepts calculated ib the double-logarithmic approximation (DLA).
The weakest point of Refs.~\cite{ber1},\cite{ber} is the fact that
they keep $\alpha_s$  fixed (at some unknown scale). It leads therefore to
the intercepts $\Delta^{DL}, \Delta^{DL}_{NS}$  explicitly
depending on this unknown coupling,
whereas $\alpha_s$  is well-known to be running. The results of
Refs.~\cite{ber1},\cite{ber}  led many authors
(see e.g.\cite{kw}) to suggest that the DGLAP parametrization
$\alpha_s = \alpha_s(Q^2)$ has to be used. However, according to results
of  Ref.~\cite{egt1}, such a parametrization is
correct only for $x \sim 1$ and
cannot be used for $x \ll 1$. The explicit dependence of  $\alpha_s$
suggested in Ref.~\cite{egt1} has been used to calculate both $g_1^{NS}$
and $g_1$ at small $x$ in Refs.~\cite{egt2}. The present talk is based on
the results obtained in those papers.

Instead of a direct study of $g_1$, it is more convenient to consider the
forward Compton amplitude $A$ related to $g_1$ as follows:

\begin{equation}
\label{g1a}
g_1(x, Q^2) = \frac{1}{\pi} \Im_s A(s, Q^2) ~.
\end{equation}

We cannot use DGLAP for studying $g_1$ or $A$ at small $x$ because it does
not account for double-log (and single-log)
contributions which are independent of $Q^2$. In order to
account for  the double-logs of both $x$ and  $Q^2$,
we need to construct two-dimensional
evolution equations that would combine the $x$- and $Q^2$- evolutions.
On the other hand, these equations should sum up the contributions of
the Feynman graphs involved to all orders in $\alpha_s$. Some of those
graphs have either ultraviolet or infrared (IR) divergencies. The
ultraviolet divergencies are regulated by the usual renormalization
procedure. In order  to regulate the IR ones, we have to introduce an IR
cut-off.
We use the IR cut-off $\mu$ in the transverse momentum space for
momenta $k_i$ of all virtual quarks and gluons:

\begin{equation}
\label{mu}
\mu < k_{i \perp}
\end{equation}
where $k_{i \perp}$ stands for the transverse (with respect to the plane
formed by the external momenta $p$ and $q$) component of $k_i$. This way
of regulating the IR divergencies was suggested by L.N. Lipatov and used
first in Ref.~\cite{kl} for quark-quark scattering. Using this cut-off
$\mu$, $A$ acquires a dependence on $\mu$. Therefore, one can
evolve $A$ with respect to $\mu$, constructing thereby some Infrared
Evolution Equations (IREE). As $A = A(s/\mu^2, Q^2/\mu^2)$,

\begin{equation}
\label{lhs}
- \mu^2 \partial A / \partial \mu^2 =
 \partial A/ \partial \rho + \partial A/ \partial y
\end{equation}
where $\rho = \ln(s/\mu^2)$ and $y = \ln(Q^2/\mu^2)$. Eq.~(\ref{lhs}) is
the lhs of the IREE for A.
In order to write the rhs of the IREE, it is convenient to use the
Sommerfeld-Watson transform
\begin{equation}
\label{mellin}
 A(s, Q^2) =
\int_{- \imath \infty}^{\imath \infty}\frac{d \omega}{2 \pi \imath}
(s/ \mu^2)^{\omega}
\xi(\omega)  F(\omega, Q^2)
\end{equation}
where $\xi(\omega)$ is the negative signature factor,
$\xi(\omega) =  [1 - e^{ - \imath \pi \omega}]/ 2 \approx
 \imath \pi \omega / 2$.
It must be noted that  the transform inverse to Eq.~(\ref{mellin})
involves the imaginary parts of $A$:

\begin{equation}
\label{invmellin}
F(\omega, Q^2) =  \frac{2}{\pi \omega}\int_0^{\infty} d \rho
e^{- \rho \omega}
\Im A(s, Q^2) ~.
\end{equation}
Notice that contrary to the amplitude $A$, the structure function $g_1$
does not have any signature and therefore  $\xi(\omega) = 1$ when
the transform ~(\ref{mellin}) is applied directly to $g_1$.

\section{Infrared evolution equations for $g_1$}

When the factorization depicted in Fig.~\ref{dubnafig1} is assumed, 
the calculation
of $g_1$
(we will not use the superscript ``s''
for$g_1$ singlet, though we use the notaion $g_1^{NS}$ for the
non-singlet $g_1$ ) is
reduced to calculating the Feynman graphs contributing  the partonic
tensor $\tilde{W}_{\mu \nu}^s$ depicted as the upper blobs
in Fig.~\ref{dubnafig1}. 
Both cases, (a), when the virtual photon scatters off the
nearly on-shell polarized quark, and (b), when the quark is replaced by
the polarized gluon, should be taken into account. Therefore, in contrast
to Eq.~(\ref{g1a}), we need to introduce two Compton amplitudes: $A_q$
and $A_g$ corresponding to the upper blob in 
Fig.~\ref{dubnafig1}a and Fig.~\ref{dubnafig1}b
respectively. The subscripts ``q'' and ``g''
refer to the initial partons. Therefore,

\begin{equation}
\label{g1qg}
g_1(x, Q^2) = g_q(x, Q^2) + g_q(x, Q^2),
\end{equation}
where
\begin{equation}
\label{gqg}
~g_q = \frac{1}{\pi} \Im_s A_q(s, Q^2),
~g_g = \frac{1}{\pi} \Im_s A_g(s, Q^2) ~
\end{equation}

\begin{figure}
\begin{center}
\begin{picture}(280,160)
\put(10,10){
\epsfbox{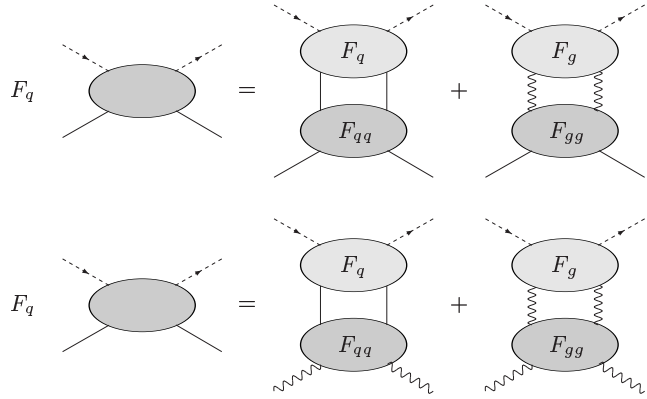}
}
\end{picture}
\end{center}
\caption{Infrared evolution equations for the amplitudes $A_q, A_g$.}
\label{dubnafig2}
\end{figure}

Let us now construct the IREE for
the amplitudes $A_{q,g}$ related to $g_1$.
To this aim, let us consider a virtual parton
with minimal $k_{\perp}$. We call such a parton the softest one.
If it is a gluon, its DL contribution can be
factorized, i.e. its DL contribution comes from the graphs where
its propagator is attached to the external lines. As the gluon propagator
cannot be attached to photons, this case is absent in IREE for $A_{q,g}$.
The second option is when the softest partons are a $t$-channel
quark-antiquark or gluon pair. It leads us to the IREE depicted in
Fig.~\ref{dubnafig2}.
Applying the operator $- \mu^2 \partial / \partial \mu^2$ to it,
combining the result with Eq.~(\ref{lhs}) and
using (\ref{mellin}), we arrive at the following system of equations:

\begin{eqnarray}
\label{system}
\big( \omega + \frac{\partial}{\partial y}\big) F_q(\omega, y) =
\frac{1}{8 \pi^2} \big[ F_{qq}(\omega) F_q(\omega, y)  +
F_{qg}(\omega) F_g (\omega, y)\big] ~, \nonumber \\
\big( \omega + \frac{\partial}{\partial y}\big) F_g(\omega, y) =
\frac{1}{8 \pi^2} \big[ F_{gq}(\omega) F_q(\omega, y)  + F_{gg}(\omega)
 F_g(\omega, y) \big] ~.
\end{eqnarray}
The amplitudes
$F_q, F_g$ are related to  $A_q, A_g$ through the transform
(\ref{mellin}).
The Mellin amplitudes $F_{ik}$, with $i,k = q,g$, describe the
parton-parton forward scattering. They contain DL contributions to
all orders in $\alpha_s$. We can introduce the new anomalous
dimensions $H_{ik} = (1/8\pi^2)F_{ik}$. The way of writing the
subscripts ``q,g'' corresponds to
the DGLAP-notations.
Solving this system of eqs. and using Eq.~(\ref{gqg}) leads to

\begin{eqnarray}
\label{gqgg}
g_q(x, Q^2) = \int_{- \imath \infty}^{\imath \infty}
\frac{d \omega}{2 \pi \imath} (1/ x)^{\omega}
\Big[C_+(\omega) e^{\Omega_+ y} +
C_-(\omega) e^{\Omega_- y} \Big] ~, \\ \nonumber
g_g(x, Q^2) = \int_{- \imath \infty}^{\imath \infty}
\frac{d \omega}{2 \pi \imath} (1/ x)^{\omega}
\Big[C_+(\omega) \frac{X + \sqrt{R}}{2H_{qg}} e^{\Omega_+ y} +
C_-(\omega) \frac{X - \sqrt{R}}{2H_{qg}} e^{\Omega_- y} \Big]
\end{eqnarray}

The unknown factors $C_{\pm}(\omega)$
have to be specified. All other factors in   Eq.~(\ref{gqgg}) can be
expressed in therms of $H_{ik}$:

\begin{eqnarray}
\label{xromega}
X = H_{gg} - H_{qq} ,~
R = (H_{gg} - H_{qq})^2 +4 H_{qg}H_{gq} ~ \\ \nonumber
\Omega_{\pm} =
\frac{1}{2}\left[H_{qq} + H_{gg} \pm
\sqrt{(H_{qq} - H_{gg})^2 + 4H_{qg}H_{gq}} \right] .
\end{eqnarray}

The anomalous dimension matrix $H_{ik}$ was calculated in Ref.~\cite{egt2}:

\begin{eqnarray}
\label{solh}
H_{gg} &=&\frac{1}{2} \Big(\omega + Y + \frac{ b_{qq} - b_{gg}}{Y}\Big),\qquad
H_{qq}= \frac{1}{2} \Big(\omega + Y - \frac{ b_{qq} - b_{gg}}{Y}\Big),
\\ \nonumber
H_{gq} &=& -\frac{b_{gq}}{Y},\qquad H_{qg} = -\frac{b_{qg}}{Y} ~.
\end{eqnarray}
where

\begin{equation}
\label{Y}
Y = - \sqrt{ \Big(\omega^2 \!-\! 2( b_{qq} \!+\! b_{gg}) \!+\!
\sqrt{\left[(\omega^2 \!-\! 2( b_{qq} \!+\! b_{gg}))^2  \!-\! 
4(b_{qq} \!-\! b_{gg})^2
\!-\! 16 b_{qg} b_{gq} \right]} ~ \Big)/2} ~,
\end{equation}
\begin{equation}
\label{b}
b_{ik} = a_{ik} + V_{ik} ,
\end{equation}
\begin{equation}
\label{aik}
a_{qq} = \frac{A(\omega)C_F}{2\pi},
~a_{gg} = \frac{2A(\omega) N}{\pi},~
a_{gq} = -\frac{n_f A'(\omega)}{2\pi},~
a_{qg} =\frac{ A'(\omega)C_F}{\pi} ~,
\end{equation}
and
\begin{equation}
\label{vborn}
V_{ik} =\frac{ m_{ik}}{\pi^2}D(\omega) ,
\end{equation}
with
\begin{equation}
\label{m}
m_{qq} =  \frac{C_F}{2N},~
m_{gg} = -2N^2,~
m_{qg} = n_f \frac{N}{2},~
m_{gq} = -N C_F.
\end{equation}
We have used here the notations $C_F = 4/3, N = 3$ and $n_f = 4$.
The factors $A$ and $D$ account for running
$\alpha_s$. They are given by the following
expressions:

\begin{equation}
\label{a}
A(\omega) = \frac{1}{b} \Big[\frac{\eta}{\eta^2 + \pi^2} -
\int_{0}^{\infty} \frac{d \rho e^{- \omega \rho}}
{(\rho + \eta)^2 + \pi^2}\Big] ~,
\end{equation}
\begin{equation}
\label{d}
D(\omega) = \frac{1}{2b^2} \int_0^{\infty} d \rho e^{- \omega \rho}
\ln \big( (\rho + \eta)/ \eta\big)
\Big[  \frac{\rho + \eta}{(\rho + \eta)^2 + \pi^2} +
\frac{1}{\rho + \eta}\Big]
\end{equation}
with $\eta = \ln(\mu^2/ \Lambda_{QCD}^2)$ and $b = (33 - 2 n_f)/ 12 \pi$.
$A'$ is defined as  $A$ with the $\pi^2$ term dropped out.
Now we can specify the coefficients $C_{\pm}$ of  Eq.~(\ref{gqgg}). When
$Q^2 = \mu^2$,
 \begin{equation}
\label{match}
g_q = \tilde{\Delta} q(x_0), ~~~~~~~~g_g = \tilde{\Delta} g(x_0)
\end{equation}
where $ \tilde{\Delta} q(x_0)$ and $ \tilde{\Delta} g(x_0)$ are
the input distributions of the polarized partons at $x_0 = \mu^2/s$.
They do not depend on $Q^2$.
Using Eq.~(\ref{match}) allows to express $C_{\pm}(\omega)$ in terms of
 $\Delta q(\omega)$ and $\Delta g(\omega)$,
which are related to $\tilde{\Delta} q(x_0)$
and $\tilde{\Delta} g(x_0)$ through the ordinary Mellin transform. Indeed,

\begin{equation}
\label{matchomega}
C_+ + C_- = \Delta q,
~~~~C_+\frac{X + \sqrt{R}}{2H_{qg}} +
C_- \frac{X - \sqrt{R}}{2H_{qg}} = \Delta g ~ .
\end{equation}
This leads to the following expressions for $g_q$ and $g_g$:
\begin{equation}
\label{gqsol}
g_q(x, Q^2) = \int_{- \imath \infty}^{\imath \infty}
\frac{d \omega}{2 \pi \imath} (1/ x)^{\omega}
\Big[ \Big(  A^{(-)} \Delta q +
B \Delta g \Big)  e^{\Omega_+ y} +
 \Big( A^{(+)} \Delta q - B \Delta g \big) e^{\Omega_- y} \big]~,
\end{equation}

\begin{equation}
\label{ggsol}
g_g(x, Q^2) = \int_{- \imath \infty}^{\imath \infty}
\frac{d \omega}{2 \pi \imath} (1/ x)^{\omega}
\Big[\Big( E \Delta q + A^{(+)}\Delta g
\Big) e^{\Omega_+ y} +
\Big(- E\Delta q +  A^{(-)}\Delta g \Big)
 e^{\Omega_- y}\Big]
\end{equation}
with
\begin{equation}
\label{abe}
A^{(\pm)} = \Big(\frac{1}{2} \pm \frac{X}{2 \sqrt{R}}\Big),~
B = \frac{H_{qg}}{\sqrt{R}},~
E = \frac{H_{gq}}{\sqrt{R}} ~.
\end{equation}

Eqs.~(\ref{gqsol}, (\ref{ggsol}) express $g_1$ in terms of the
parton distributions $\Delta q(\omega)$ and $\Delta g(\omega)$. However,
they are related to the distributions
  $\tilde{\Delta} q(x_0)$ and $\tilde{\Delta} g(x_0)$ at very low $x$:
 $x_0 \approx \mu^2/s \ll 1$. Therefore, they scarcely
can be found from experimental data. It is much more useful to
express $g_q, g_g$ in terms of the initial parton densities
 $\tilde{\delta} q$ and $\tilde{\delta} g$
defined at $x \sim 1$. We can do it, using the evolution of
  $\tilde{\Delta} q(x_0)$,~ $\tilde{\Delta} g(x_0)$ with respect to
$s$. Indeed, the $s$-evolution of $\tilde{\delta} q, \tilde{\delta} q$
from $s \approx \mu^2$ to $s \gg \mu^2$
 at fixed $Q^2$ ~$(Q^2 = \mu^2)$ is equivalent to their $x$-evolution
from $x \sim 1$ to $x \ll~1$. In the $\omega$-space, the system of IREE
for the parton distributions looks quite similar to  Eqs.~(\ref{system}).
However, the eqs for $\Delta q,~ \Delta g$ are algebraic because they
do not depend on $Q^2$:

\begin{eqnarray}
\label{systeminput}
  \Delta q (\omega) &=& (<e^2_q>/2) \delta q(\omega) + (1 /\omega)
\left[H_{qq}(\omega) \Delta q(\omega)  +
H_{qg}(\omega) \Delta g (\omega) \right]~, \nonumber \\
 \Delta g(\omega) &=& (<e^2_q>/2)\hat{\delta g}(\omega) + (1/\omega)
\left[H_{gq}(\omega) \Delta q(\omega)  +
H_{gg}(\omega) \Delta g(\omega) \right] ~.
\end{eqnarray}
 where $<e^2_q>$ is the sum of the quark electric
charges ($<e^2_q> = 10/9$ for $n_f = 4$),  $\delta q$ is the sum of the 
initial quark and antiquark densities  and  
$\hat{\delta g} \equiv  -  (A'(\omega)/2 \pi \omega^2) \delta g$
is the starting
point of the evolution of the gluon density $\delta g$. It corresponds to
Fig.~\ref{dubnafig1}b where the upper blob is substituted by the quark box.
Solving Eqs.~(\ref{systeminput}), we
obtain:

\begin{equation}
\label{inputq}
\Delta q=
(<e^2_q>/2) \frac{
\big[\omega (\omega -H_{gg}) \delta q + \omega H_{qg}\hat{\delta g}\big] }
{\big[\omega^2 - \omega(H_{qq} + H_{gg}) + (H_{qq}H_{gg} -
H_{qg}H_{gq})\big]}~,
\end{equation}
\begin{equation}
\label{inputg}
\Delta g=
(<e^2_q>/2) \frac{
\big[\omega H_{gq} \delta q + \omega(\omega - H_{qq})\hat{\delta} g \big]}
{\big[\omega^2 - \omega(H_{qq} + H_{gg}) + (H_{qq}H_{gg} -
H_{qg}H_{gq})\big]}  ~.
\end{equation}

Then Eqs.~(\ref{gqsol},\ref{ggsol},\ref{inputq},\ref{inputg}) express
$g_1$ in
terms of the initial parton densities  $\delta q, \delta g$.

When we put $H_{qg}= H_{gq}= H_{gg} = 0$ and do not sum over $e_q$, we
arrive at the expression for the non-singlet structure function $g_1^{NS}$:
Obviously, in this case $A^{(+)} = B = E = \Omega_- = 0,~
A^{(-)} = 1,~~ \Omega_+ = H_{qq}$. However, the nonsinglet
anomalous dimension $ H_{qq}$ should be  calculated in the limit
 $b_{gg} = b_{qg} = b_{gq} = 0$. We denote such  $H_{qq} \equiv H^{NS}$.
The explicit expression for it is:
 \begin{equation}
\label{f0ns}
H^{NS} =
(1/2) \Big[\omega - \sqrt{\omega^2 - 4 b_{qq}} \Big]~.
\end{equation}
Therefore, we arrive at

 \begin{equation}
\label{gns}
g_1^{NS} = \frac{e^2_q}{2} \int_{ -\imath \infty}^{\imath \infty}
\frac{d \omega}{2 \pi \imath}
\Big(\frac{\omega \delta q}{\omega - H^{NS} } \Big)
\Big( 1/x \big)^{\omega}
\Big( Q^2/\mu^2\big)^{H^{NS}  } ~.
\end{equation}

\section{Small-$x$ asymptotics for $g_1$ }

When $x \to 0$ and $Q^2 \gg \mu^2$, one can
neglect contributions with $\Omega_-$ in  Eqs.~(\ref{gqgg}). As is known,
$g_1 \sim (1/x)^{\omega_0}$ at $x \to 0$, with $\omega_0$ being the position
of the leading singularity of the integrand of $g_1$ .
According to  Eqs.~(\ref{solh}),  the leading singularity, $\omega^{NS}$
for $g_1^{NS}$ is the rightmost
root of the equation
 \begin{equation}
\label{singns}
\omega^2 - 4 b_{qq} = 0
\end{equation}
while the leading singularity,
$\omega_0$ for $g_1$ is the rithmost root of

 \begin{equation}
\label{sings}
\omega^4 - 4 (b_{qq} + b_{gg})\omega^2 + 16
(b_{qq} b_{gg} -   b_{qg} b_{gq}) = 0~.
\end{equation}

\begin{figure}
\begin{center}
\begin{picture}(320,220)
\put(0,20){
\epsfxsize=12cm
\epsfysize=3cm
\epsfbox{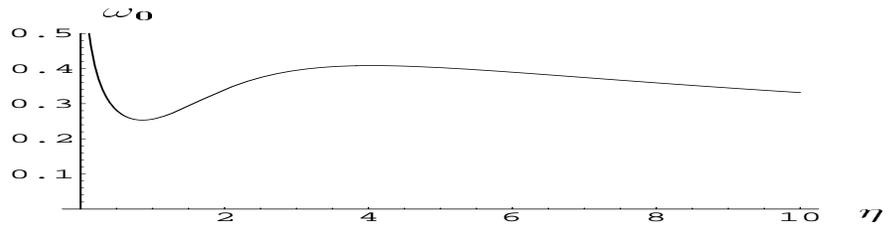}
}
\end{picture}
\end{center}
\caption{Dependence of the intercept $\omega_0$ on infrared cut-off
$\eta=\ln(\mu^2/\Lambda_{QCD})$ for $g_1^{NS}$.}
\label{dubnafig3}
\end{figure}

\begin{figure}
\begin{center}
\begin{picture}(270,180)
\put(10,10){
\epsfbox{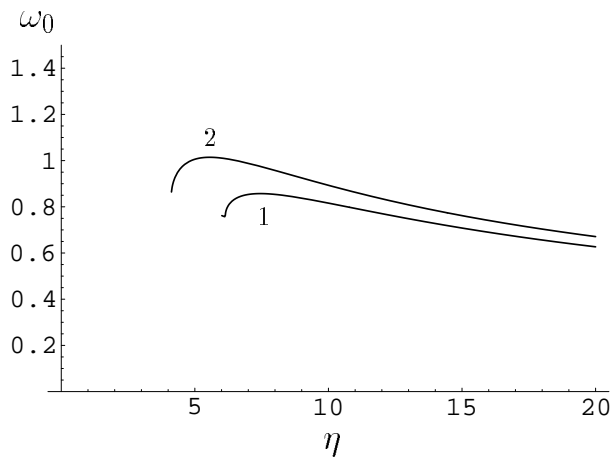}}
\end{picture}
\caption{Dependence on $\eta$ of the rightmost root of
Eq.~(\ref{sings}), $\omega_0$. Curve~2 corresponds to the case
when gluon contributions only are taken into account;
curve~1 is the result of accounting for both gluon and quark
contributions.}
\label{dubnafig4}
\end{center}
\end{figure}

In our approach, all factors $b_{ik}$ depend on
$\eta = \ln(\mu^2/\Lambda_{QCD})$, so the roots of
Eqs.~(\ref{singns},\ref{sings}) also depend on $\eta$.
This dependence is plotted in Fig.~\ref{dubnafig3} for $\omega^{NS}$  and in
Fig.~\ref{dubnafig4} for $\omega_0$.
Both the curve in Fig.~\ref{dubnafig3} and the curve~1 in 
Fig.~\ref{dubnafig4} have
a maximum. We denote this maximum as the intercept. Therefore,
\begin{equation}
\label{g1asympt}
g_1^{NS} \sim (e^2_q/2) \delta q
\left(\frac{1}{x}\right)^{\Delta_{NS}}
\left(\frac{Q^2}{\mu^2}\right)^{\Delta_{NS}/2},\qquad
g_1 \sim (1/2) [Z_1 \delta q + Z_2 \delta_g]
\left(\frac{1}{x}\right)^{\Delta_S}
\left(\frac{Q^2}{\mu^2}\right)^{\Delta_S/2},
\end{equation}
and we find for the intercepts
\begin{equation}
\label{intercepts}
\Delta_{NS} \approx 0.4,\qquad \Delta_S \approx 0.86
\end{equation}
and
$Z_1 = - 1.2,\quad Z_2 = -0.08$.
This implies that $g_1^{NS}$ is positive when $x \to 0$ whereas $g_1$
can be either positive or negative, depending on the relation between
$\delta q$ and $\delta g$. In particular, $g_1$ is positive when
\begin{equation}
\label{posit}
15\,\delta q + \delta g < 0 .
\end{equation}
otherwise it is negative. In other words, the sign of $g_1$ at small $x$
can be positive if the initial gluon density is negative and large.

\section{Conclusion}
\label{CONCLUSION}

The total resummation of the most singular $(\sim \alpha^n_s/\omega^{2n +
1})$ terms in the expressions for the anomalous dimensions and the
coefficient
functions leads to
the expressions of
Eqs.~(\ref{g1qg},\ref{gqsol},\ref{ggsol},\ref{gns}) for the
singlet and the non-singlet structure functions $g_1$.  It
garantees the Regge
(power-like) behaviour (\ref{g1asympt}) of $g_1,~ g_1^{NS}$ when $x \to 0$,
with the intercepts given by  Eq.~(\ref{intercepts}). The intercepts
$\Delta_{NS}, \Delta_S$ are of course obtained with the running QCD
coupling effects taken into account. The value of
the non-singlet intercept $\Delta_{NS} = 0.4$ is now
confirmed by several
independent analyses
\cite{kat} of experimental data.
The value  $\Delta_S = 0.86$ of the singlet intercept is in a good
agreement with the estimate  $\Delta_S = 0.88 \pm 0.14$ obtained in
Ref.~\cite{koch} from analysis of the HERMES data.

Another interesting point
to discuss is the sign of these structure functions.
Eq.~(\ref{gns}) states
that $g_1^{NS}$ is positive both at $x \sim 1$ and at $ x \ll 1$.
The situation concerning the singlet $g_1$ is more involved: being
positive at
$x \sim 1$, the singlet $g_1$ can remain positive at $x \ll 1$
only if the initial
parton densities obey Eq.~(\ref{posit}), otherwise it becomes negative.

\section{Acknowledgement}
The work is supported by grants POCTI/FNU/49523/2002
and RSGSS-1124.2003.2  One of us (B.I Ermolaev) is grateful to the Organizing
Commettee of the Workshop for their hospitality.


\begin{thebibliography}{99}
\bibitem{dglap} G.~Altarelli and G.~Parisi. Nucl.Phys.B 126(1977) 298;\\
 V.N.~Gribov and L.N.~Lipatov. Sov.J.Nucl.Phys.15(1978) 438 and 675;\\
 Yu.L.~Dokshitzer.  Sov.Phys.JETP 46(1977)641.
\bibitem{a} G.~Altarelli, R.~Ball, S.~Forte and G.~Ridolfi.
Acta Phys.Polon.B29(1998)1201, Nucl.Phys.B496(1999)337.
\bibitem{ber1} B.I.~Ermolaev, S.I.~Manaenkov and M.G.~Ryskin.
Z.Phys.C69(1996)259;\\ ~J.~Bartels, B.I.~Ermolaev and M.G.~Ryskin.
Z.Phys.C 70(1996)273.
\bibitem{ber} J.~Bartels, B.I.~Ermolaev and M.G.~Ryskin.
Z.Phys.C 72(1996)627.
\bibitem{kw} J.~Kwiecinski. Acta.Phys.Polon.B 29(2001)1201;\\
D.~Kot\-lorz and A.~Kot\-lorz. Acta.Phys.Polon.B 32(2001)2883,\\
B.~Badalek, J.~Kiryluk and J.~Kwiecinski.
Phys.Rev.D 61(2000)014009;\\
J.~Kwiecinski and B.~Ziaja. Phys.Rev.D 60(1999)9802386;\\
B.~Ziaja. Phys. Rev. D66(2002)114017; hep-ph/0304268.
\bibitem{egt1} B.I.~Ermolaev, M.~Greco and S.I.~Troyan.
Phys.Lett.B~522(2001)57.
\bibitem{egt2} B.I.~Ermolaev, M.~Greco and S.I.~Troyan.
Nucl.Phys.B~594(2001)71; ibid 571(2000)137; hep-ph/0307128.
\bibitem{kl} R.~Kirschner and L.N.~Lipatov. Sov.Phys.JETP 56(1982)266.
\bibitem{ggfl} V.N.~Gribov, V.G.~Gorshkov, G.V.~Frolov, L.N.~Lipatov.
Sov.J.Nucl.Phys. 6(1968)95, ibid 6(1968)262.
\bibitem{kat} J.~Soffer and O.V.~Teryaev. Phys.Rev.D 56(1997)1549;
A.L.~Kataev, G.~Parente, A.V.Sidorov. CERN-TH-2001-058,
Phys.Part.Nucl.34(2003)20,\\ Fiz.Elem.Chast.Atom.Yadra 34(2003)43,
Nucl.Phys.A666(2000)184;\\
A.V.~Kotikov, A.V.~Lipatov, G.~Parente, N.P.~Zotov, Eur.Phys.J.C26(2002)51;
V.G.~Krivokhijine, A.V.~Kotikov, hep-ph/0108224;\\
A.V.~Kotikov, D.V~. Peshekhonov,  hep-ph/0110229.
\bibitem{koch}   N.I.~Kochelev, K.~Lipka, W.-D.~Nowak, V.~Vento,
A.V.~Vinnikov.  Phys.Rev. D67 (2003) 074014.
\end{thebibliography}
\end{document}